\documentclass[prd,twocolumn,showpacs,amsmath,amssymb,nofootinbib,superscriptaddress]{revtex4}
\usepackage{graphicx}
\usepackage[pdftex]{color}
\usepackage{bm}
\usepackage{amsfonts}
\usepackage{amssymb}
\usepackage{color}
\usepackage{float}
\usepackage{amsmath}
\usepackage{dcolumn}
\usepackage{hyperref}
\usepackage{amsfonts}

\renewcommand{\[}{\left[}
\renewcommand{\]}{\right]}

\def\nn{\nonumber}

\def\be{\begin{equation}}
\def\ee{\end{equation}}
\def\bea{\begin{eqnarray}}
\def\eea{\end{eqnarray}}
\def\eqi{\begin{equation}}
\def\eqf{\end{equation}}
\def\eqia{\begin{eqnarray}}
\def\eqfa{\end{eqnarray}}

\newcommand{\fs}{{\rm{\it f\sigma_8}}}
\begin{document}

\title{Cosmological constraints on $\gamma$-gravity models}

\author{Clara \'{A}lvarez Luna}\email{c.a.luna@ucm.es}
\affiliation{Departamento de F\'{\i}sica Te\'{o}rica, Universidad Complutense de Madrid, Plaza de las Ciencias 1, 28040 Madrid, Spain}

\author{Spyros Basilakos}\email{svasil@academyofathens.gr}
\affiliation{Academy of Athens, Research Center for Astronomy and
Applied Mathematics, Soranou Efesiou 4, 11527 Athens, Greece}

\author{Savvas Nesseris}\email{savvas.nesseris@csic.es}
\affiliation{Instituto de F\'isica Te\'orica UAM-CSIC, Universidad Auton\'oma de Madrid, Cantoblanco, 28049 Madrid, Spain}

\date{\today}

\begin{abstract}
In this paper we place observational constraints on the well-known $\gamma$-gravity $f(R)$ model using the latest cosmological data, namely we use the latest growth rate, Cosmic Microwave Background, Baryon Acoustic Oscillations, Supernovae type Ia and Hubble parameter data. Performing a joint likelihood analysis we find that the $\gamma$-gravity model is in very good agreement with observations. Utilizing the AIC statistical test we statistically compare the current $f(R)$ model with $\Lambda$CDM cosmology and find that they are statistically equivalent. Therefore, $\gamma$-gravity can be seen as a useful scenario toward testing deviations from General Relativity. Finally, we note that we find somewhat higher values for the $f(R)$ best-fit values compared to those  mentioned in the past in the literature and we argue that this could potential alleviate the halo-mass function problem.
\end{abstract}

\maketitle

\section{Introduction}
Over the last two decades a plethora of independent cosmological studies (see Ref.~\cite{Ade:2015xua} and references therein) have shown that the observed Universe is spatially flat to a very high precision. Furthermore, the energy budget of the Universe is roughly $\sim 30\%$ dark and baryonic matter, while the rest $\sim 70\%$ is the so-called dark energy (DE). The latter component is frequently used in order to understand the accelerated expansion of the Universe, despite the fact that the related underlying microscopic physics is still unknown.

Recently, a large number of viable cosmological models have been proposed in the literature, providing different
physical explanations regarding the cosmic acceleration. These models can be separated into two general groups.
The first family of models adheres to General Relativity (GR) and it introduces new fields in nature, see for example Refs.~\cite{Copeland:2006wr,Caldwell:2009ix}. The second possibility is to build modified gravity models, which have GR as a particular limit, but with additional degrees of freedom that can cause acceleration~\cite{Copeland:2006wr,Clifton:2011jh}.

Of course the usual forms of matter (dark matter, baryonic and radiation) cannot provide an explanation for the accelerated expansion of the Universe, since the corresponding equation-of-state (EoS) parameter $w$ is always positive or equal to zero. Considering that any of the usual kinds of matter behaves as an ideal fluid, then the EoS can be defined as the ratio of the fluid's pressure $P$ to its density $\rho$, namely $w=\frac{P}{\rho}$. In the case of non-relativistic matter the EoS parameter is strictly equal to zero, $w_{m}=0$. Concerning the radiation component it is well known that the EoS is given by $P_{r}=\frac13 \rho_{r}$, hence $w_{r}=\frac13$. On the other hand, several dark energy models (quintessence and the like) have $w \ge -1$, where the cosmological constant is recovered for $w=-1$. Moreover, in the case of modified gravity models the EoS parameter $w$ can cross the phantom line, namely $w<-1$, which would normally be attributed to ghost fields~\cite{Nesseris:2006er}.

However, it is not enough to study the expansion history, but also to see the effect on the evolution of perturbations, as the latter are responsible for the Large Scale Structure. The effect of dark energy on the growth of perturbations is therefore an important tool in discriminating models from $\Lambda$CDM, see Refs.~ \cite{Bertschinger:2006aw,Blake:2012pj,Knox:2005rg,Laszlo:2007td,Nesseris:2007pa}.

The case of $f(R)$ models occupies  an  eminent position  in  the  hierarchy  of modified gravity models. Based on a simple extension of the Einstein-Hilbert action, $f(R)$ models appear  to  be  ideal  tools  for studying cosmic acceleration, testing theories of structure formation  and  extracting  invaluable  cosmological  information. For example, Starobinsky~\cite{Starobinsky:1980te} was the first who introduced the simple Lagrangian $f(R)=R+m R^2$ toward studying early inflation. Nevertheless, it has been found~\cite{Amendola:2006kh,Amendola:2006we} that some types of $f(R)$ models are unable to provide the correct matter era thus they are not viable.

On the other hand, it has been shown that a large body of $f(R)$ models can support a proper matter dominated era, see Refs.~\cite{Nojiri:2006gh,Capozziello:2006dj,Odintsov:2017hbk,Elizalde:2010ts}, which is essential for structure formation. For more details regarding the properties of $f(R)$ models we refer the reader to the reviews of Refs.~\cite{DeFelice:2010aj,Sotiriou:2008rp,Nojiri:2010wj,Nojiri:2017ncd}. Moreover in  Refs.~\cite{Tsujikawa:2007gd,Tsujikawa:2007tg}, one may find a detailed discussion concerning the evolution of matter perturbations and the related Newton's parameter $G_{\textrm{eff}}$. Lastly, observational constraints on $f(R)$ models can be found in Refs.~\cite{Basilakos:2013nfa,Abebe:2013zua,Dossett:2014oia,delaCruz-Dombriz:2015tye}, while some methods which reconstruct $f(R)$ models from observations can be seen in Refs.~\cite{Lee:2017lud,Lee:2017dox,Perez-Romero:2017njc}.

One of the most popular and viable $f(R)$ model which contains a proper matter era and passes the solar system tests is the so called $\gamma$-gravity model~\cite{ODwyer:2013vfo}. This model has a very steep dependence on the Ricci scalar $R$, hence it is in agreement with the expected results from large scale structure~\cite{ODwyer:2013vfo}. Moreover, the $\gamma$-gravity model has been used in N-body simulations~\cite{Santos:2016vdv}, where it has been found that, for some ad-hoc choices of the model parameters, there were significant deviations between the predicted halo mass function with that of observations. Specifically, the authors of Ref.~\cite{Santos:2016vdv} considered values of the $\alpha$ parameter of the theory to be in the range $\alpha \in \[1.05,1.5\]$, while as we will present later current observations prefer a much higher value, namely $\alpha\simeq 1.9\pm0.2$.

Clearly, even though there is no formal correspondence of the $\gamma$-gravity model to GR, the steepness of the gamma functions guarantee that strong deviations on the Large Scale Structure (LSS) can not only be suppressed but also adjusted via the parameter $\alpha$ \cite{ODwyer:2013vfo}. Compared to other similarly viable $f(R)$ models, this steep dependence of the gamma-gravity model on the Ricci scalar can help distinguish gamma-gravity from other alternatives, something which was  already demonstrated by the authors of Ref. \cite{ODwyer:2013vfo}. On the other hand, for the same reason, the effect on the growth of perturbations can be significant, see for example Fig 8 in Ref. \cite{ODwyer:2013vfo}. Regarding the dynamics of the model, it was shown in Ref.~\cite{Santos:2016vdv}, see their Fig. 1, that this model can indeed mimic $\Lambda$CDM to a few percent in the equation of state $w(z)$ at late times and almost perfectly for higher redshifts, thus allowing for the possibility to successfully discriminate these two models.

Furthermore, even though this model is interesting enough that it was used to perform N-body simulations in Ref.~\cite{Santos:2016vdv}, as mentioned we found that the values of the $\alpha$ parameter used in that analysis were too low with respect to the ones actually preferred by the observational data, thus possibly leading to potential biases in the interpretations of the simulations. Both of these points provide enough motivation to study gamma-gravity models in more detail, as we have done in this paper.

In our work we implement a joint likelihood analysis in order to put strong constraints on the $\gamma$-gravity model, involving data form (JLA) type Ia supernovae, Planck 2015 CMB shift parameters, Baryon Acoustic Oscillation and ``Gold 2017" growth rate. The layout of our manuscript is as follows: in Sec.~\ref{backevo} we provide the basic elements of $f(R)$ gravity, in Sec.~\ref{funcforms} we present the $f(R)$ $\gamma$-gravity model, while in Sec.~\ref{constrs} we provide the observational constraints on the fitted model parameters and we compare the $\gamma$-gravity and the $\Lambda$CDM models. Finally, we discuss our conclusions in Sec.~\ref{conclusions}.

\section{$f(R)$ gravity and cosmology \label{backevo}}
In this section we briefly present $f(R)$ gravity in the context of Friedmann$-$Lemaitre$-$Robertson$-$Walker (FLRW) metric. Specifically, we consider a homogeneous,  isotropic and spatially flat universe
that contains non-relativistic matter (baryon) and radiation. The modified Einstein-Hilbert action is given by
\be
S=\int d^{4}x\sqrt{-g}\left[  \frac{1}{2\kappa^{2}}f\left(  R\right)
+\mathcal{L}_{m}+\mathcal{L}_{r}\right],  \label{action1}%
\ee
where $\mathcal{L}_{m,r}$ are the matter and radiation Lagrangians and $\kappa^{2}=8\pi G_N$.
If we vary the action of Eq.~(\ref{action1}) with respect to the metric then we obtain the corresponding field equations:
\bea
F G_{\mu\nu}&-&\frac12(f(R)-R~F) g_{\mu\nu}+\left(g_{\mu\nu}\Box-\nabla_\mu\nabla_\nu\right)F\nn \\ &=&\kappa^{2}\,T_{\mu\nu},
\label{EE}
\eea
where $F=f'(R)$, $T_{\mu\nu}$ is the total energy-momentum tensor for all species and $G_{\mu\nu}$ is the Einstein tensor.

In the framework of a spatially flat FLRW with Cartesian coordinates
\be
ds^{2}=-dt^{2}+a^{2}(t) d\vec{x}^{2} \label{metric},
\ee
 the modified Friedmann equations can be written as
\bea
3 F H^{2}&-&\frac{F R-f}{2}+3H\dot{F} =\kappa^{2}(\rho_{m}+\rho_{r}), \label{fried1}\\%
-2F\dot{H}&=&\kappa^{2}\left(\rho_{m}+\frac43\rho_{r}\right) +\ddot{F}-H\dot{F},\label{fried2}%
\eea
where $F_{R}=\partial_R F=f''(R)$, $a(t)=\frac{1}{1+z}$ is the scale factor
of the universe,
while the time derivative of the Ricci scalar is given by
$\dot{R}=aH\frac{dR}{da}$.
Also, the Ricci scalar in this case takes the form
\begin{equation}
R=g^{\mu\nu}R_{\mu\nu}= 6\left(  \frac{\ddot{a}}{a}+\frac{\dot
{a}^{2}}{a^{2}}\right)  =6(2H^{2}+\dot{H}) \;.\label{ricci}
\end{equation}

We observe that solving analytically the system of Eqs.~(\ref{fried1}) and~(\ref{fried2}) is in general not possible, hence we need to solve it numerically. In this kind of studies it is useful to introduce
the effective (``geometrical'') dark energy EoS parameter in terms of $E(a)=H(a)/H_{0}$~\cite{Huterer:2000mj,Nesseris:2004wj}
\begin{equation}
\label{eos222}
w(a)=\frac{-1-\frac{2}{3}a\frac{{d\rm lnE}}{da}}
{1-\Omega_{m}(a)},
\end{equation}
where we have set
\be
\label{ddomm}
\Omega_{m}(a)=\frac{\Omega_{m0}a^{-3}}{E^{2}(a)} \;.
\ee
Obviously, in the case of $f(R)=R-2\Lambda$ the current equations
reduce to those of $\Lambda$CDM model, namely
the EoS parameter is exactly $-1$ and the Hubble parameter is given by
\be
H_{\Lambda}(a)^2/H_0^2=\Omega_{m0}a^{-3}+\Omega_{r0}a^{-4}+1-\Omega_{m0}-\Omega_{r0} \;\;.\label{Hlcdm}
\ee

Lastly, we would like to remind the reader that in the context of $f(R)$ gravity the Newton's parameter $G_{\rm eff}$ is affected by the scale and redshift (or scale factor). In particular, at sub-horizon scales we have
(see~\cite{Tsujikawa:2007gd,Tsujikawa:2007tg}):
\be
\frac{G_{\textrm{eff}}}{G_N}=\frac1{F}\frac{1+4\frac{k^2}{a^2}\frac{F_{,R}}{F}}{1+3\frac{k^2}{a^2}\frac{F_{,R}}{F}}\label{geff},
\ee
where of course $G_N$ is Newton's constant and $k$ is the relevant scale of the Fourier modes. As expected for $\Lambda$CDM model we find $G_{\textrm{eff}}/G_N=1$.

Notice, that we restrict our analysis to the choice of $k=300H_0$ which implies $k\sim0.1 h/\textrm{Mpc}$. The reason for this is that, while in general the growth rate in $f(R)$ models will be scale-dependent so we should not restrict our analysis to just one value of $k$, in practice we are interested in the perturbations that are relevant to the linear regime of galaxy clusterings and are accessible to the surveys. This corresponds to the wave numbers $0.01 h/\textrm{Mpc} < k < 0.1 h/\textrm{Mpc}$ or equivalently $30 < k/H_0 < 300$, see Ref.~\cite{Tegmark:2006az}. In the end we choose the value of $k=300 H_0=0.1h/\textrm{Mpc}$ as that roughly corresponds to the limit before we enter the non-linear regime.

Furthermore, in order to have a viable $f(R)$ model the corresponding of Newton's parameter $G_{\textrm{eff}}$ needs to satisfy some important conditions. Specifically, the first one is $G_{\textrm{eff}}>0$ due to the fact that gravitons should always carry positive energy. Second, Big Bang Nucleosynthesis poses the following condition $G_{\textrm{eff}}/G_N=1.09\pm 0.2$~\cite{Bambi:2005fi}. Third, the ratio $G_{\textrm{eff}}(a=1)/G_N$ is normalized to unity at the present time. Finally, the functional form of $G_{\rm eff}$ affects the evolution of linear matter perturbations $\delta=\frac{\delta\rho_m}{\rho_m}$, via the following differential equation~\cite{Tsujikawa:2007gd}
\be
\ddot{\delta}+2 H \dot{\delta}=4 \pi G_{\textrm{eff}} \delta \label{growthode}.
\ee
However, in order to directly compare with observations we need to introduce the parameter $f(a)\sigma_{8}(a)$, where $f(a)=\frac{d \log \delta}{d \log a}$ is the growth rate of clustering and $\sigma_8(a)=\sigma_{8,0}\frac{\delta(a)}{\delta(1)}$ is the root mean square (rms) fluctuations on $8h^{-1}$Mpc. Combining the latter expressions we obtain the relation $f\sigma_8(a)=\sigma_{8,0} \frac{\delta'(a)}{\delta(1)}$
which is used toward fitting the growth rate data. The reader may find more details in Ref.~\cite{Nesseris:2017vor} and references therein.

\section{The $f(R)$ $\gamma$-gravity functional form and the numerical approach\label{funcforms}}

\begin{figure}[!t]
	\centering
	\includegraphics[width=0.5\textwidth]{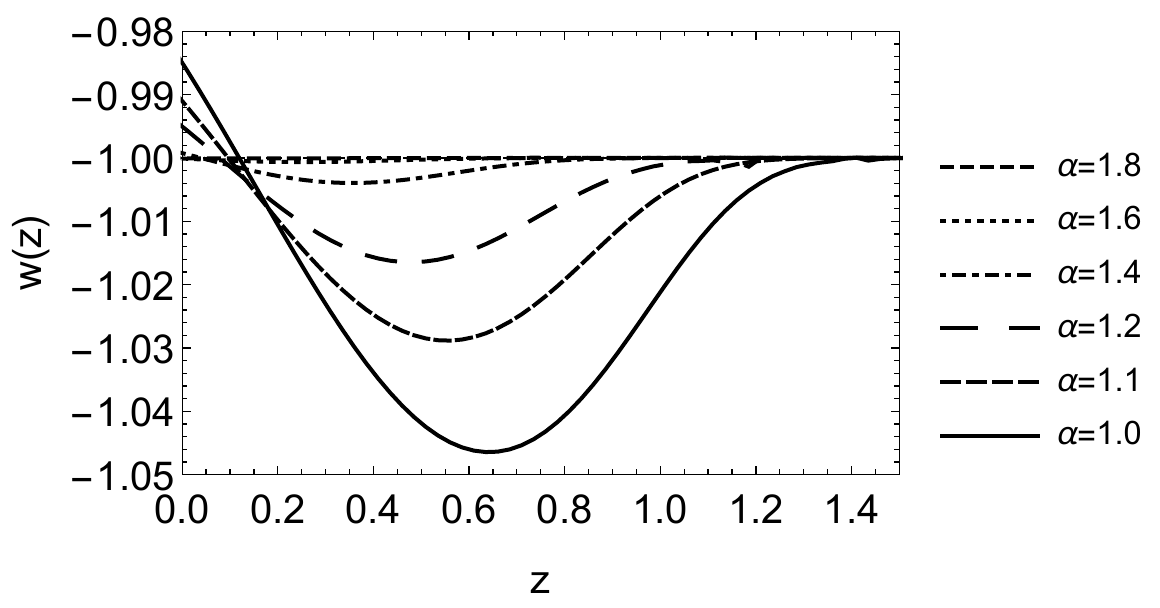}
	\caption[]{The effective equation of state $w_{\rm DE}(z)$ of the $\gamma-$gravity model for $n=2$, $\Omega_{m0}=0.28$ and $\alpha$=(1,1.1,1.2,1.4,1.6,1.8). As can be seen, for realistic values of the parameter $\alpha$ the deviation of the background expansion from the $\Lambda$CDM model can be significant, for example it can reach approximately $\sim5\%$ at intermediate redshifts. }
	\label{fig:wz}
\end{figure}

In this section we briefly present the basic ingredients of the so called $f(R)$ $\gamma-$gravity model. Here the
Lagrangian of the model is $f(R)=R+\widetilde{f}(R)$, where
the function $\widetilde{f}(R)$ is given by:
\be
\widetilde{f}(R)=-\frac{\alpha R_s}{n} \gamma\left(\frac{1}{n},\left(\frac{R}{R_s}\right)^n\right).\label{eq:fR}
\ee
In the latter formula $\alpha, R_s$ are constants, $n$ is a positive integer and $\gamma(n,z)=\int_0^z t^{n-1} e^{-t}dt$ is the incomplete $\Gamma$-function.
Evidently, for
$n=1$ the $\gamma-$gravity model boils down to the case of exponential gravity first discussed in Ref.~\cite{Cognola:2007zu}, but see also Ref.~\cite{Linder:2009jz}. In this work following the notations of~\cite{Santos:2016vdv} we utilize $n=2$ and thus we can define
$m^2=\Omega_{m0} H_0^2$, $d=\frac{1-\Omega_{m0}}{\Omega_{m0}}$ and
\be
\frac{R_s}{m^2}=\frac{6\;n\;d}{\alpha \Gamma(1/n)} .
\ee
Then, using the variable $N=\ln(a)$ the first modified
Friedmann equation becomes:
\be
H(N)^2(1+\widetilde{f}_R+R'(N)\widetilde{f}_{RR})-\frac{R \widetilde{f}_R-\widetilde{f}}{6}=m^2 \exp(-3N),\label{eq:fried}
\ee
where we have set $\widetilde{f}_R=\widetilde{f}'(R)$ and $\widetilde{f}_{RR}=\widetilde{f}''(R)$ with $\widetilde{f}(R)$ given by Eq.~(\ref{eq:fR}). Notice that the Ricci scalar is given by $R=12H(N)^2+6 H(N)H'(N)$. If we introduce the new variables
\bea
x_1(N)&=&\frac{H^2}{m^2}-e^{-3N}-d,\\
x_2(N)&=&\frac{R}{m^2}-3e^{-3N}-12 (d+x_1),
\eea
then the main cosmological equations take the following
simple forms
\bea
x_1'(N)&=&\frac{x_2}{3}, \label{eq:odex1}\\
x_2'(N)&=&\frac{R'}{m^2}+9e^{-3N}-4x_2.\label{eq:odex2}
\eea
Obviously, the latter system can be easily solved numerically.
Lastly, in the context of $(x_{1},x_{2})$ variables
the effective EoS parameter can be written as
\be
w_{\rm DE}=-1-\frac19 \frac{x_2}{x_1+d}.
\ee
The $\gamma$-gravity model is rather interesting since in principle can be distinguished from $\Lambda$CDM. As can be seen in Fig. \ref{fig:wz}, the EoS $w(z)$ can deviate from -1, by more than $5\%$ at intermediate redshifts for realistic values of $\alpha$, something which could be detectable by future surveys like Euclid and LSST as they claim they will be able to make per cent measurements of the equation of state $w(z)$.

To actually compare the $\gamma$-gravity model to the data we need to solve the ordinary  differential equations (ODE) presented earlier numerically. However, due to the stiffness of the system of and the fact that at high redshifts $\gamma$-gravity exactly emulates $\Lambda$CDM, we first solve numerically the equations for the background given by Eqs.~(\ref{eq:odex1})-(\ref{eq:odex2}) with initial conditions that correspond to $\Lambda$CDM at early times (around redshift  $z\sim 2$) and then patch the solution to $\Lambda$CDM all the way to recombination. Then, we also solve the ODE for the growth via Eq.~(\ref{growthode}) with initial conditions corresponding to matter domination ($\delta(a) \sim a$ and $\delta'(a) \sim 1$) at $z\sim 100$.

\begin{table}[!t]
\caption{The $H(z)$ data used in the current analysis (in units of $\textrm{km}~\textrm{s}^{-1} \textrm{Mpc}^{-1}$). This compilation is based partly in those of Refs.~\cite{Moresco:2016mzx} and \cite{Guo:2015gpa}.}
\label{tab:Hzdata}
\small
\centering
\begin{tabular}{cccccccccc}
\\
\hline\hline
$z$  & $H(z)$ & $\sigma_{H}$ & Ref.   \\
\hline
$0.07$    & $69.0$   & $19.6$  & \cite{Zhang:2012mp}  \\
$0.09$    & $69.0$   & $12.0$  & \cite{STERN:2009EP} \\
$0.12$    & $68.6$   & $26.2$  & \cite{Zhang:2012mp}  \\
$0.17$    & $83.0$   & $8.0$   & \cite{STERN:2009EP}    \\
$0.179$   & $75.0$   & $4.0$   & \cite{MORESCO:2012JH}   \\
$0.199$   & $75.0$   & $5.0$   & \cite{MORESCO:2012JH}   \\
$0.2$     & $72.9$   & $29.6$  & \cite{Zhang:2012mp}   \\
$0.27$    & $77.0$   & $14.0$  & \cite{STERN:2009EP}   \\
$0.28$    & $88.8$   & $36.6$  & \cite{Zhang:2012mp}  \\
$0.35$    & $82.7$   & $8.4$   & \cite{Chuang:2012qt}   \\
$0.352$   & $83.0$   & $14.0$  & \cite{MORESCO:2012JH}   \\
$0.3802$  & $83.0$   & $13.5$  & \cite{Moresco:2016mzx}   \\
$0.4$     & $95.0$   & $17.0$  & \cite{STERN:2009EP}    \\
$0.4004$  & $77.0$   & $10.2$  & \cite{Moresco:2016mzx}   \\
$0.4247$  & $87.1$   & $11.2$  & \cite{Moresco:2016mzx}   \\
$0.44$    & $82.6$   & $7.8$   & \cite{Blake:2012pj}   \\
$0.44497$ & $92.8$   & $12.9$  & \cite{Moresco:2016mzx}   \\
$0.4783$  & $80.9$   & $9.0$   & \cite{Moresco:2016mzx}   \\
$0.48$    & $97.0$   & $62.0$  & \cite{STERN:2009EP}   \\
$0.57$    & $96.8$   & $3.4$   & \cite{Anderson:2013zyy}   \\
$0.593$   & $104.0$  & $13.0$  & \cite{MORESCO:2012JH}  \\
$0.60$    & $87.9$   & $6.1$   & \cite{Blake:2012pj}   \\
$0.68$    & $92.0$   & $8.0$   & \cite{MORESCO:2012JH}    \\
$0.73$    & $97.3$   & $7.0$   & \cite{Blake:2012pj}   \\
$0.781$   & $105.0$  & $12.0$  & \cite{MORESCO:2012JH} \\
$0.875$   & $125.0$  & $17.0$  & \cite{MORESCO:2012JH} \\
$0.88$    & $90.0$   & $40.0$  & \cite{STERN:2009EP}   \\
$0.9$     & $117.0$  & $23.0$  & \cite{STERN:2009EP}   \\
$1.037$   & $154.0$  & $20.0$  & \cite{MORESCO:2012JH} \\
$1.3$     & $168.0$  & $17.0$  & \cite{STERN:2009EP}   \\
$1.363$   & $160.0$  & $33.6$  & \cite{Moresco:2015cya}  \\
$1.43$    & $177.0$  & $18.0$  & \cite{STERN:2009EP}   \\
$1.53$    & $140.0$  & $14.0$  & \cite{STERN:2009EP}  \\
$1.75$    & $202.0$  & $40.0$  & \cite{STERN:2009EP}  \\
$1.965$   & $186.5$  & $50.4$  & \cite{Moresco:2015cya}  \\
$2.34$    & $222.0$  & $7.0$   & \cite{Delubac:2014aqe}   \\
\hline\hline
\end{tabular}
\end{table}

\begin{table*}[t!]
\caption{The ``Gold-2017" compilation of $f\sigma_8(z)$ measurements from different surveys, compiled in Ref.~\cite{Nesseris:2017vor}. In the columns we show in ascending order with respect to redshift, the name and year of the survey that made the measurement, the redshift and value of $f\sigma_8(z)$ and the corresponding reference and fiducial cosmology. These data points are used in our analysis in the next sections.
\label{tab:fsigma8data}}
\begin{centering}
\begin{tabular}{ccccccc}
Index & Dataset & $z$ & $f\sigma_8(z)$ & Refs. & Year & Notes \\
\hline
1 & 6dFGS+SnIa & $0.02$ & $0.428\pm 0.0465$ & \cite{Huterer:2016uyq} & 2016 & $(\Omega_m,h,\sigma_8)=(0.3,0.683,0.8)$ \\

2 & SnIa+IRAS &0.02& $0.398 \pm 0.065$ &  \cite{Turnbull:2011ty},\cite{Hudson:2012gt} & 2011& $(\Omega_m,\Omega_K)=(0.3,0)$\\

3 & 2MASS &0.02& $0.314 \pm 0.048$ &  \cite{Davis:2010sw},\cite{Hudson:2012gt} & 2010& $(\Omega_m,\Omega_K)=(0.266,0)$ \\

4 & SDSS-veloc & $0.10$ & $0.370\pm 0.130$ & \cite{Feix:2015dla}  &2015 &$(\Omega_m,\Omega_K)=(0.3,0)$ \\

5 & SDSS-MGS & $0.15$ & $0.490\pm0.145$ & \cite{Howlett:2014opa} & 2014& $(\Omega_m,h,\sigma_8)=(0.31,0.67,0.83)$ \\

6 & 2dFGRS & $0.17$ & $0.510\pm 0.060$ & \cite{Song:2008qt}  & 2009& $(\Omega_m,\Omega_K)=(0.3,0)$ \\

7 & GAMA & $0.18$ & $0.360\pm 0.090$ & \cite{Blake:2013nif}  & 2013& $(\Omega_m,\Omega_K)=(0.27,0)$ \\

8 & GAMA & $0.38$ & $0.440\pm 0.060$ & \cite{Blake:2013nif}  & 2013& \\

9 &SDSS-LRG-200 & $0.25$ & $0.3512\pm 0.0583$ & \cite{Samushia:2011cs} & 2011& $(\Omega_m,\Omega_K)=(0.25,0)$  \\

10 &SDSS-LRG-200 & $0.37$ & $0.4602\pm 0.0378$ & \cite{Samushia:2011cs} & 2011& \\

11 &BOSS-LOWZ& $0.32$ & $0.384\pm 0.095$ & \cite{Sanchez:2013tga}  &2013 & $(\Omega_m,\Omega_K)=(0.274,0)$ \\

12 & SDSS-CMASS & $0.59$ & $0.488\pm 0.060$ & \cite{Chuang:2013wga} &2013& $\ \ (\Omega_m,h,\sigma_8)=(0.307115,0.6777,0.8288)$ \\

13 &WiggleZ & $0.44$ & $0.413\pm 0.080$ & \cite{Blake:2012pj} & 2012&$(\Omega_m,h)=(0.27,0.71)$ \\

14 &WiggleZ & $0.60$ & $0.390\pm 0.063$ & \cite{Blake:2012pj} & 2012& \\

15 &WiggleZ & $0.73$ & $0.437\pm 0.072$ & \cite{Blake:2012pj} & 2012 &\\

16 &Vipers PDR-2& $0.60$ & $0.550\pm 0.120$ & \cite{Pezzotta:2016gbo} & 2016& $(\Omega_m,\Omega_b)=(0.3,0.045)$ \\

17 &Vipers PDR-2& $0.86$ & $0.400\pm 0.110$ & \cite{Pezzotta:2016gbo} & 2016&\\

18 &FastSound& $1.40$ & $0.482\pm 0.116$ & \cite{Okumura:2015lvp}  & 2015& $(\Omega_m,\Omega_K)=(0.270,0)$\\
\end{tabular}\par\end{centering}
\end{table*}

\section{Observational constraints \label{constrs}}
In this section we test the statistical performance of the $\gamma$-gravity model against the latest observational data. In particular we use the JLA SnIa data of Ref.~\cite{Betoule:2014frx}, the BAO from 6dFGS\cite{Beutler:2011hx}, SDDS\cite{Anderson:2013zyy}, BOSS CMASS\cite{Xu:2012hg}, WiggleZ\cite{Blake:2012pj}, MGS\cite{Ross:2014qpa} and BOSS DR12\cite{Gil-Marin:2015nqa}. We also use the CMB shift parameters based on the \textit{Planck 2015} release \cite{Ade:2015xua}, as derived in Ref.~\cite{Wang:2015tua}.

We also assume a flat Universe and that the radiation density at the present epoch, which is relevant for the CMB shift parameter, is fixed to $\Omega_{r0}=\Omega_{m0} a_{eq}$, where the scale factor at equality is $a_{eq}=\frac{1}{1+2.5~10^4 \Omega_{m0} h^2 \left(T_{cmb}/2.7K\right)^{-4}}$. We also marginalize over the parameters $M$ and $\delta M$ of the JLA set as described in the appendix of Ref.~\cite{Conley:2011ku}. These parameters implicitly contain $H_0$ and thus the SnIa $\chi^2$ term is independent of $H_{0}=h~100km/s/Mpc$, where according to Planck $h\simeq0.67$ \cite{Ade:2015xua}. However, we keep the parameters $\alpha_{JLA}, \beta_{JLA}$ free in our analysis.

We also used the Hubble expansion $H(z)$ data, which are derived in two ways. The first is via the differential age method, i.e. the redshift drift of distant objects over a long time period since the following relation is valid in GR $H(z)=-\frac{1}{1+z}\frac{dz}{dt}$ \cite{Jimenez:2001gg}. The second approach is through the clustering of galaxies or quasars, which provide direct measurements of $H(z)$ by constraining the BAO peak in the radial direction \cite{Gaztanaga:2008xz}. The $H(z)$ data that were used in our paper are shown along with their references in Table \ref{tab:Hzdata}.

Finally, we also use the ``Gold-2017'' growth-rate data compilation by Ref.~\cite{Nesseris:2017vor} which we show present in Table~\ref{tab:fsigma8data} with their references. The growth-rate data come from measurements of redshift-space distortions, which are very important probes of large scale structure providing measurements of $\fs(a)\equiv f(a)\cdot \sigma(a)$, where $f(a)=\frac{d ln\delta}{d lna}$ is the growth rate and $\sigma(a)=\sigma_8\frac{\delta(a)}{\delta(1)}$  is  the redshift-dependent rms fluctuations of the linear density field within spheres of radius $R=8 h^{-1} \textrm{\textrm{Mpc}}$ while the parameter $\sigma_8$ is its value today. This combination is used in what follows to derive constraints for theoretical model parameters.

Measurements of $\fs(a)$ can be made by constraining the ratio of the monopole and the quadrupole multipoles of the redshift-space power spectrum which depends on $\beta=f/b$, where $f$ is the growth rate and $b$ is the bias, in a specific way defined by linear theory \cite{Percival:2008sh,Song:2008qt,Nesseris:2006er}. The combination of $f\sigma_8(a)$ is independent of bias as all bias dependence in this combination cancels out thus, it has been shown that this combination is a good discriminator of DE (Dark energy) models \cite{Song:2008qt}.

We will assume that the aforementioned datasets can be treated as statistically independent measurements. This is a rather strong statement given that for example the SnIa data and other $H(z)$ probes are sensitive to luminosity distances in the same universe and there might be spatial overlap between the two probes, so this can lead to correlations that might affect the analysis. While this is a very important point, unfortunately at the moment there is no standard way to account for it given the lack of the full correlation matrix between the different datasets, say for example the $H(z)$ and SnIa data. Thus, following standard lines, we have assumed the different datasets are uncorrelated, which is equivalent to just summing their corresponding $\chi^2$ contributions.

On another vein, measurements of the growth rate depend on clustering so they need to assume a specific model, see Refs. \cite{Hellwing:2014nma,Bose:2017myh}, which for the data used in the current analysis is $\Lambda$CDM with various values of $\Omega_{m0}$. Such model-dependent elements could affect the modified gravity analysis, as differences in the modeling between two datapoints may manifest as systematics and/or new physics. To address it, we have already implemented the correction of the Alcock-Paczynski effect as described in Ref.~\cite{Nesseris:2017vor}, that should take into account the fiducial model used in the derivation of each measurement of the growth data.

\begin{table*}[t!]
\caption{The best-fit parameters for the $\gamma$-gravity model and $\Lambda$CDM. \label{tab:bestfits}}
\begin{centering}
\begin{tabular}{cccccccc}
Model & $\alpha_{JLA}$ & $\beta_{JLA}$ & $\Omega_{m0}$ & $100\Omega_b h^2$ & $\alpha$ & $h$ &$\sigma_{8,0}$ \\\hline
$\Lambda$CDM & $0.140\pm0.007$ & $3.115\pm0.035$ & $0.314\pm0.006$ & $2.226\pm0.014$ & $-$ & $0.674\pm0.005$ &$0.744\pm0.029$\\\hline
$\gamma$-gravity & $ 0.140\pm 0.007$ & $3.113 \pm0.073$ & $0.316\pm0.006$ & $2.226\pm0.014$ & $1.892\pm0.198$ & $0.673\pm0.005$ &$0.741\pm0.029$  \\\hline
\end{tabular}
\par
\end{centering}
\end{table*}

\begin{table}[t!]
\caption{The $\chi^2$ and AIC parameters for the $\gamma$-gravity model and $\Lambda$CDM. \label{tab:chi2AIC}}
\begin{centering}
\begin{tabular}{cccc}
Model & $\chi^2$ & AIC & $\Delta$AIC \\\hline
$\Lambda$CDM & $744.901$ &$757.006$ &$0$ \\\hline
$\gamma$-gravity & $744.826$ &$758.966$ &$-1.960$ \\\hline
\end{tabular}
\par
\end{centering}
\end{table}

\begin{figure*}[!t]
\centering
\includegraphics[width = 0.33\textwidth]{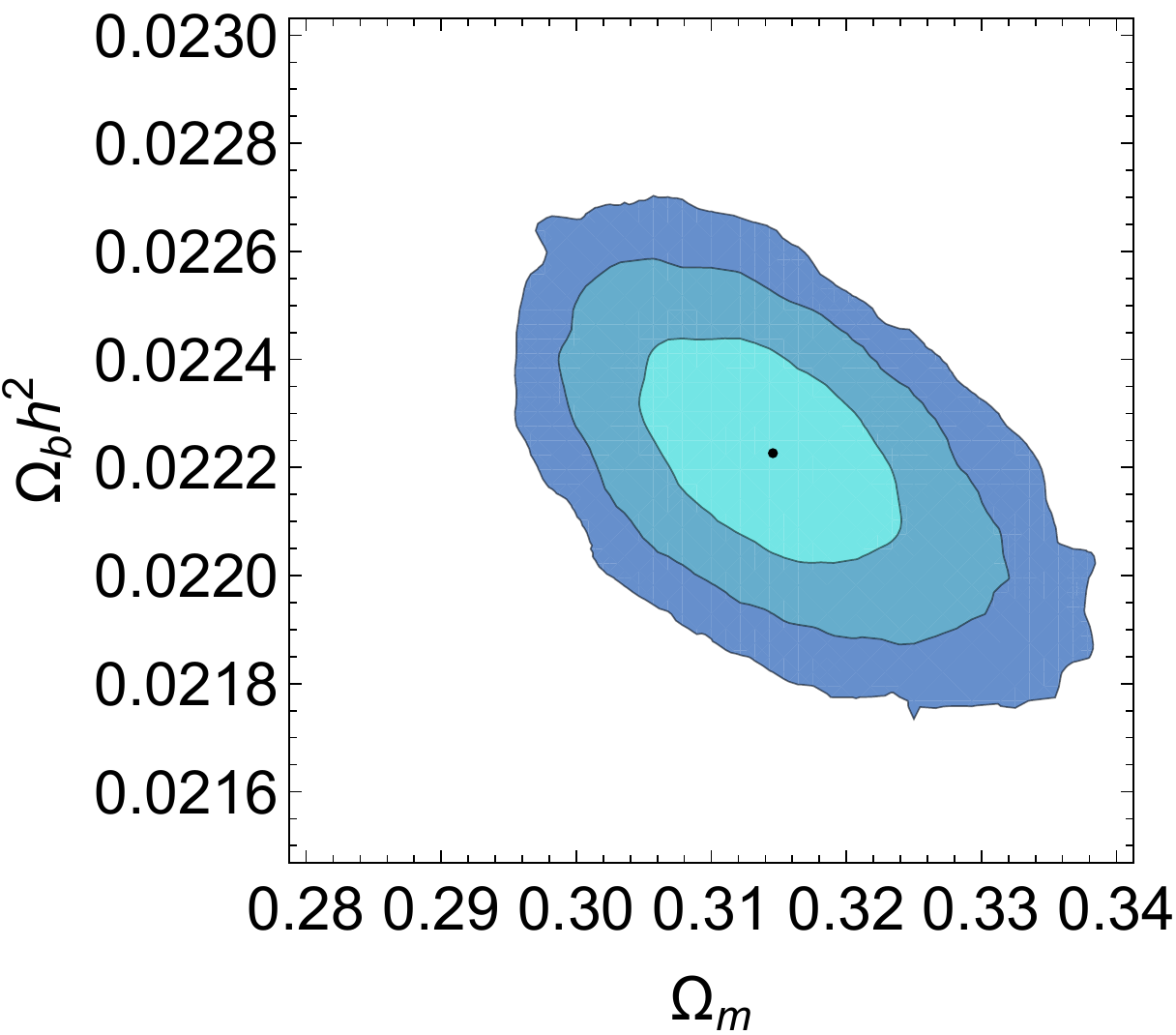}
\includegraphics[width = 0.3\textwidth]{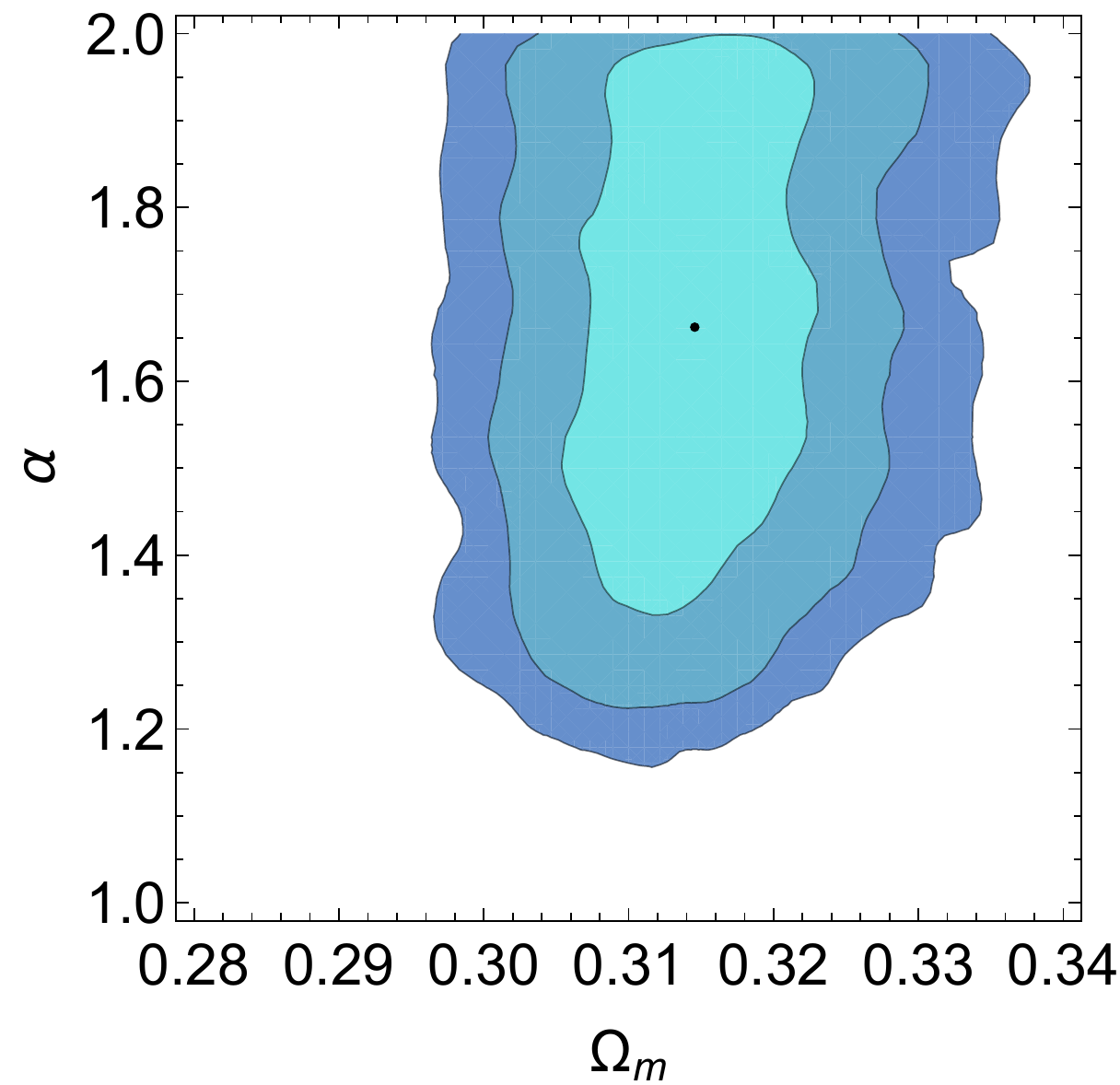}
\includegraphics[width = 0.31\textwidth]{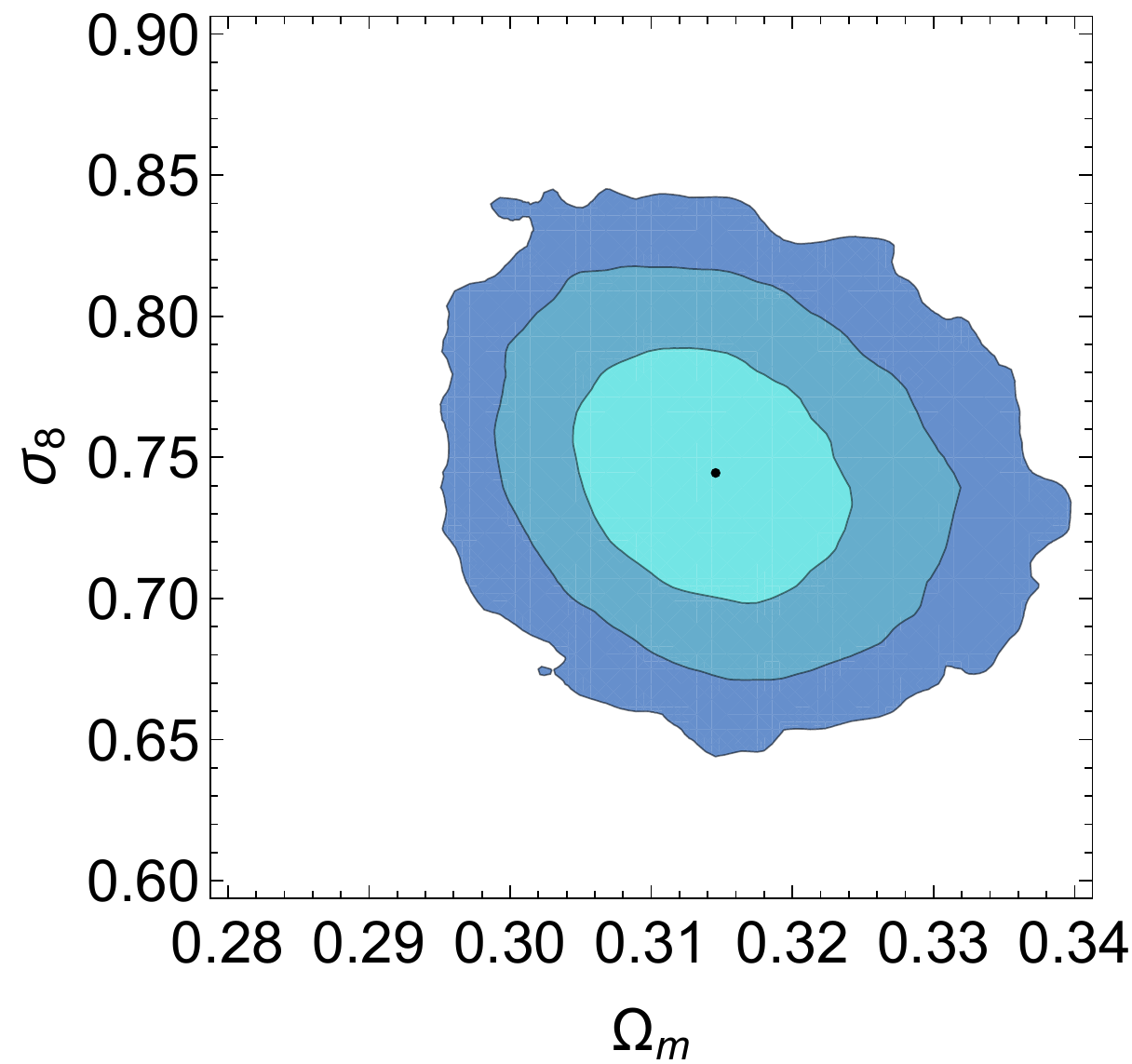}
\caption{The 68.3$\%$, 95.4$\%$ and 99.7$\%$ of the $\gamma$-gravity model for various parameter combinations. The black dots correspond to the mean values of the parameters in the MCMC chain.}
\label{fig:contoursgamma}
\end{figure*}

\begin{figure}[!t]
\centering
\includegraphics[width = 0.24\textwidth]{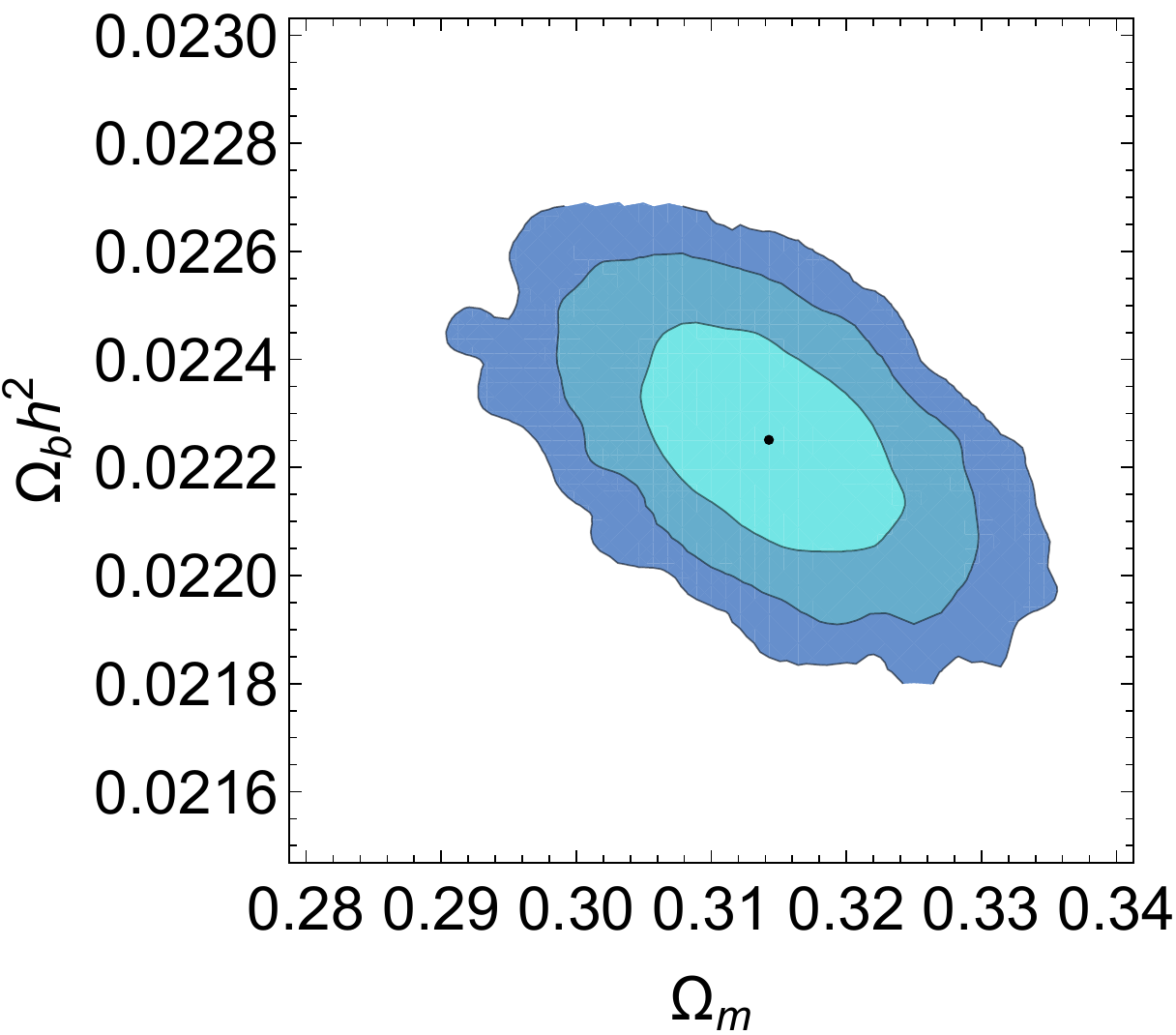}
\includegraphics[width = 0.225\textwidth]{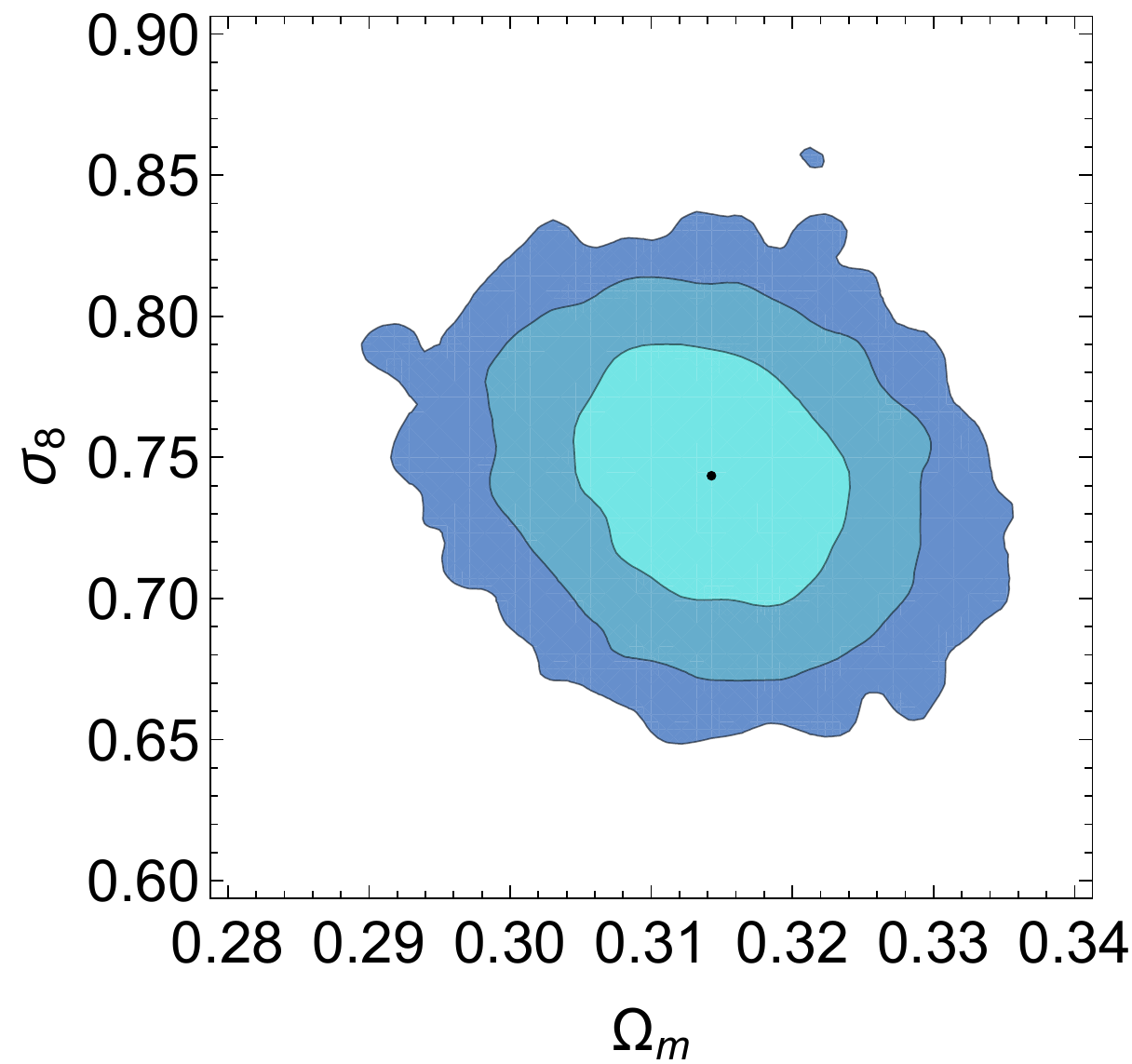}
\caption{The 68.3$\%$, 95.4$\%$ and 99.7$\%$ of the $\Lambda$CDM model for various parameter combinations. The black dots correspond to the mean values of the parameters in the MCMC chain.}
\label{fig:contoursLCDM}
\end{figure}

\begin{figure}[!t]
\centering
\includegraphics[width = 0.48\textwidth]{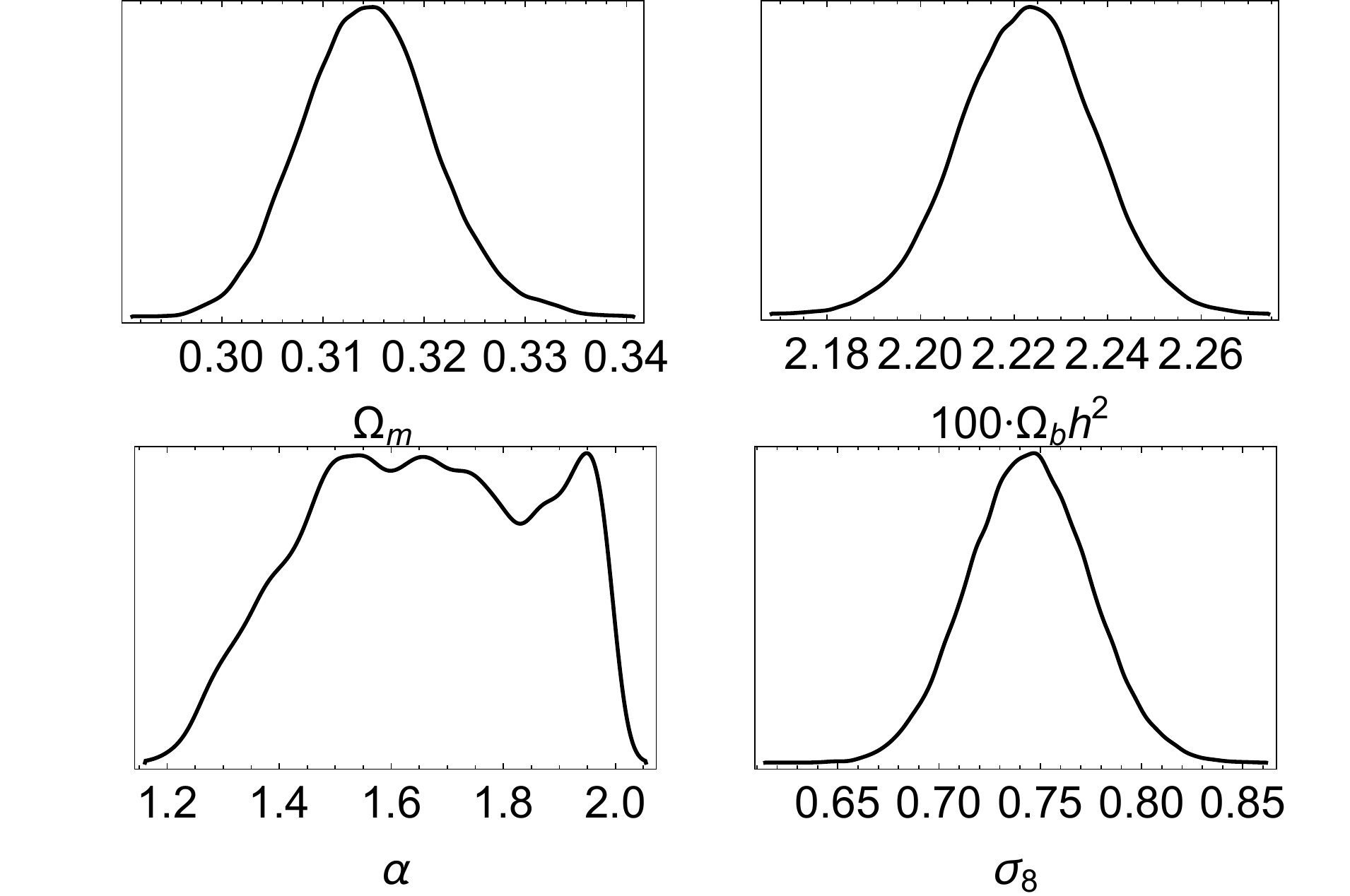}
\caption{The 1D marginalized PDFs of the $\gamma$-gravity model for various parameter combinations.}
\label{fig:1dgamma}
\end{figure}

\begin{figure*}[!t]
\centering
\includegraphics[width = \textwidth]{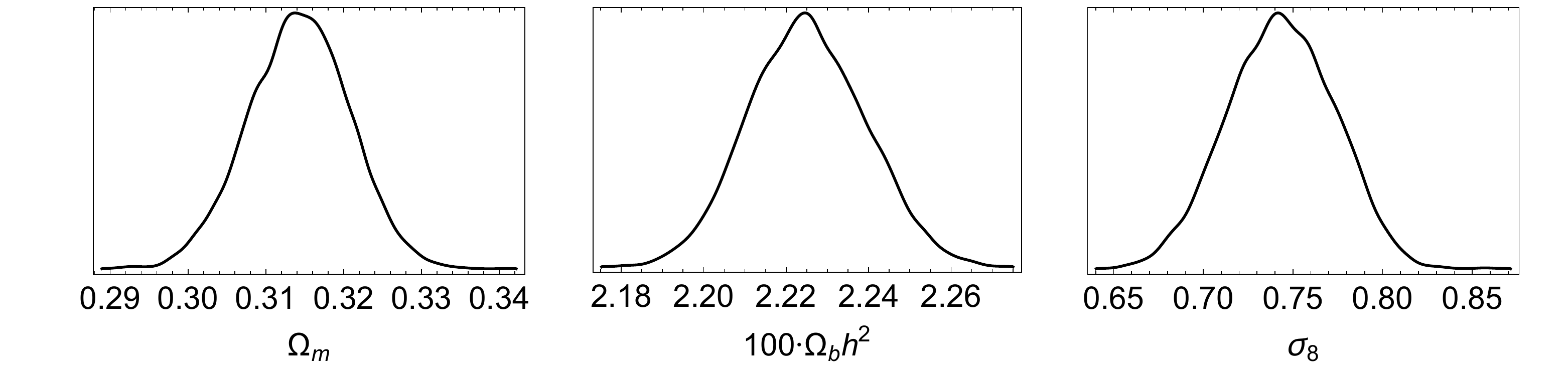}
\caption{The 1D marginalized PDFs of the $\Lambda$CDM model for $\Omega_m$ (left), $\Omega_b h^2$ (center) and $\sigma_8$ (right).}
\label{fig:1dLCDM}
\end{figure*}

\begin{figure*}[!t]
\centering
\includegraphics[width = 0.495\textwidth]{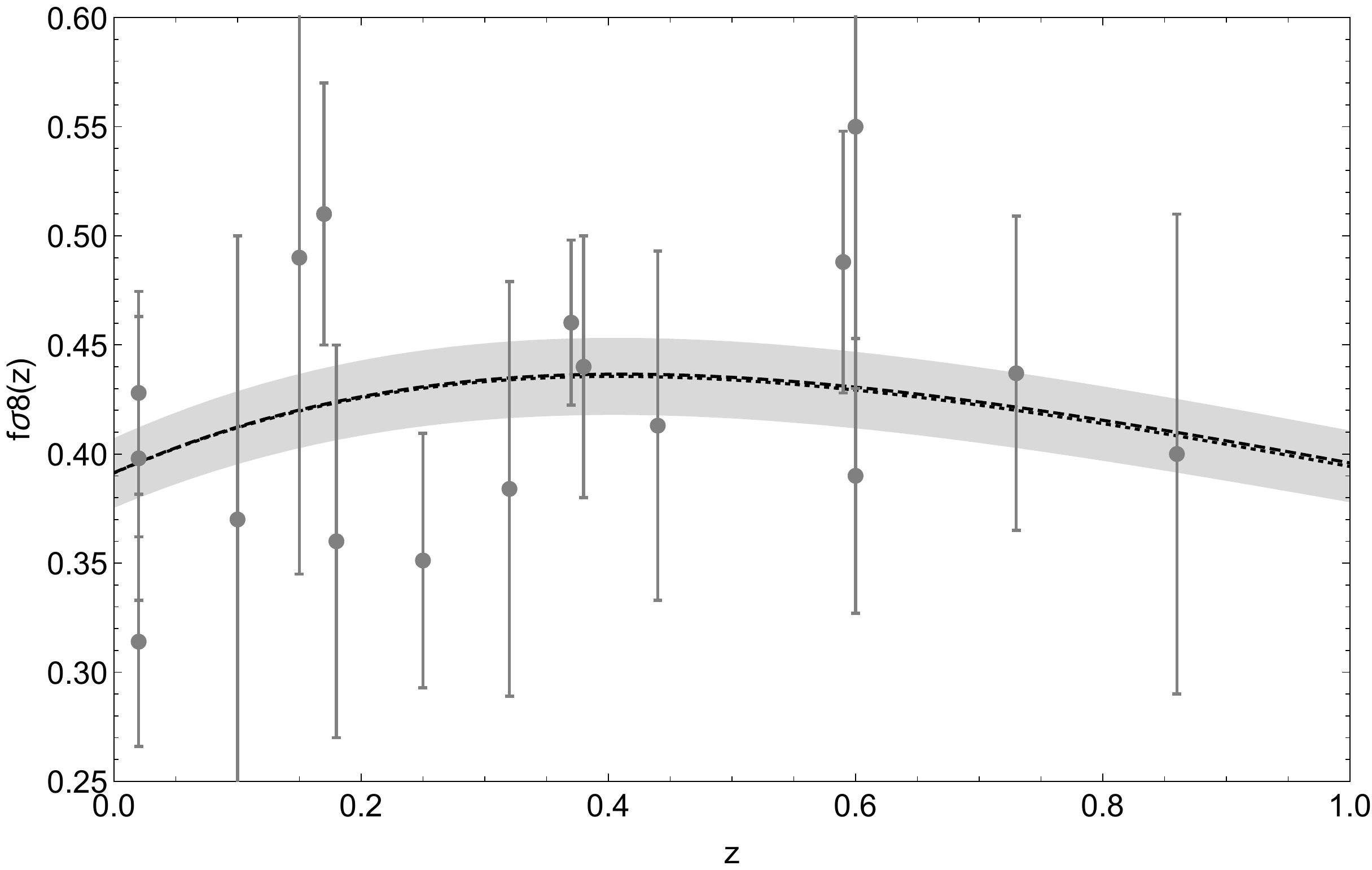}
\includegraphics[width = 0.495\textwidth]{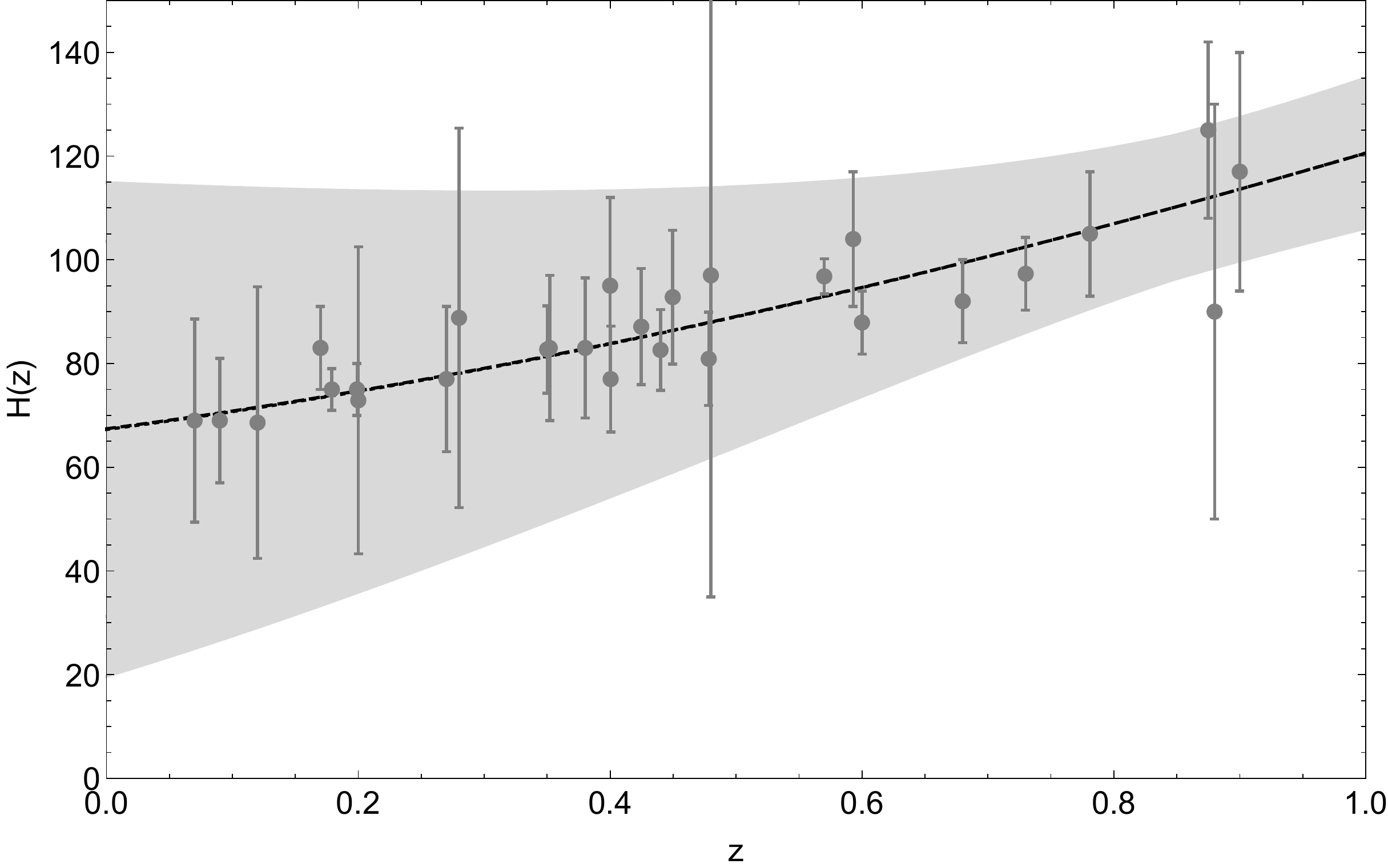}
\caption{Plots of the growth parameter $f\sigma_8(z)$ (left) and the Hubble parameter $H(z)$ (right) with their $1\sigma$ errors for the $\gamma$-gravity (dotted black line) and the $\Lambda$CDM (dashed black line) models respectively for the best-fit parameters show in Table \ref{tab:bestfits}. As can be seen, in this case the best-fit curves are practically indistinguishable from each other. }
\label{fig:plotsref}
\end{figure*}

Therefore, the overall likelihood function $L_{\rm tot}$ is given as the product of the individual likelihoods
$$
L_{\rm tot}=L_{\rm SNIa} \times L_{\rm BAO} \times L_{\rm H(z)} \times L_{\rm cmb}
\times L_{\rm gr},
$$
which means that
\be
\chi^{2}_{\rm tot}=\chi^{2}_{\rm SNIa}+\chi^{2}_{\rm BAO}+\chi^{2}_{\rm H(z)}+
\chi^{2}_{\rm cmb}+\chi^{2}_{\rm gr} .\label{eq:chi2eq}
\ee

Furthermore, we study the statistical significance of our constraints using the well known Akaike Information Criterion (AIC)~\cite{Akaike1974}. In particular, assuming Gaussian errors, the corresponding estimator is
given by
\begin{eqnarray}
{\rm AIC} = -2 \ln {\cal L}_{\rm max}+2k_p+\frac{2k_p(k_p+1)}{N_{\rm dat}-k_p-1} \label{eq:AIC}\;,
\end{eqnarray}
where $N_{\rm dat}$ and $k_p$ denote the total number of data and the number of free parameters (see also~\cite{Liddle:2007fy}). In our case we have 740 datapoints from the JLA set, the 3 CMB shift parameters, 9 BAO points, 18 growth-rate data and 36 $H(z)$ points for a total of $N_{\rm dat}=806$.

Clearly, a smaller value of AIC implies a better fit to the data. Now, if we want to compare different cosmological models then we need to introduce the model pair difference which is written as $\Delta {\rm  AIC}={\rm AIC}_{\rm model}-{\rm AIC}_{\rm min}$. The relative difference provides the following situations: $4<\Delta {\rm AIC} <7$ indicate a positive evidence against the model with higher value of ${\rm AIC}_{\rm model}$, while $\Delta {\rm AIC} \ge 10$ suggests strong evidence. On the other hand, $\Delta {\rm AIC} \le 2$ is an indication that the two comparison models are consistent.

In order to perform our statistical analysis, we consider the $\chi^2$ as given by Eq.~(\ref{eq:chi2eq}) and the parameter vectors (assuming a flat Universe) given by $p_{\Lambda \textrm{CDM}}=\left(\alpha_{JLA}, \beta_{JLA}, \Omega_{m0}, 100\Omega_b h^2, h, \sigma_{8,0}\right)$ and $p_{\gamma}=\left(\alpha_{JLA}, \beta_{JLA}, \Omega_{m0}, 100\Omega_b h^2, \alpha, h, \sigma_{8,0}\right)$ for the $\Lambda \textrm{CDM}$ and $\gamma$-gravity models respectively. Then, the best-fit parameters and their uncertainties were obtained with the aid of the MCMC method based on a Metropolis-Hastings algorithm\footnote{The MCMC code for Mathematica used in the analysis is freely available at \url{http://members.ift.uam-csic.es/savvas.nesseris/} . }. Further more, we assumed priors for the parameters given by $ \alpha_{JLA} \in[0.01, 0.2]$, $\beta_{JLA} \in[2, 4]$, $ \Omega_{m0} \in[0.1, 0.5]$, $\Omega_b h^2 \in[0.001, 0.08]$, $\alpha \in[1, 2]$, $h \in[0.4, 1]$, $\sigma_{8,0}\in[0.1, 1.8]$ and obtained $\sim10^5$ points with the MCMC code for each model.

The results of our MCMC analysis are shown in Tables~\ref{tab:bestfits} and~\ref{tab:chi2AIC} respectively. We observe that the $\gamma$-gravity model is in excellent agreement with the data. Although the $\gamma$-gravity model with $n=2$ does not include $\Lambda$CDM as a limiting case, we find that statistically it is equally competitive to $\Lambda$CDM as $\Delta \textrm{AIC} \sim 2$. This is interesting as the majority of viable $f(R)$ models, like the Hu-Sawicki \cite{Hu:2007nk} and the Starobinsky \cite{Starobinsky:2007hu}, can be seen as perturbations around $\Lambda$CDM as it was found in Ref.~\cite{Basilakos:2013nfa}. Therefore, $\gamma$-gravity can be viewed as a useful scenario toward testing deviations from GR, especially in the light of the next generation of surveys. Lastly, in Figs.~\ref{fig:contoursgamma} and~\ref{fig:contoursLCDM} we plot the 68.3$\%$, 95.4$\%$ and 99.7$\%$ contours for both models. In this context, in Figs.~\ref{fig:1dgamma} and~\ref{fig:1dLCDM} we present the 1D marginalized probability density functions (PDFs) for various parameter combinations.

In Fig.~\ref{fig:plotsref} we show the plots of the growth parameter $f\sigma_8(z)$ (left) and the Hubble parameter $H(z)$ (right) with their $1\sigma$ errors for the $\gamma$-gravity (dotted black line) and the $\Lambda$CDM (dashed black line) models respectively for the best-fit parameters show in Table \ref{tab:bestfits}. As can be seen, in this case the best-fit curves are practically indistinguishable from each other.

While part of the data we use only constrains the background cosmology, the effect of the growth data is also significant in discriminated models from $\Lambda$CDM, see Refs.~ \cite{Bertschinger:2006aw,Blake:2012pj,Knox:2005rg,Laszlo:2007td,Nesseris:2007pa}. For example, in our analysis this can be seen by extracting the contribution of the latter on the total $\chi^2$ of the two models. The individual $\chi^2$ values for the CMB, BAO, SnIa, growth and $H(z)$ datasets for both models, corresponding to the individual values for the background and growth dataset chi-squared values from our analysis, are as follows respectively:
\bea
\chi^2_{\Lambda \textrm{CDM}} &=& (0.112, 13.440, 695.435, 12.977, 22.937), \nn \\
\chi^2_{\gamma} &=& (0.293, 12.864, 695.519, 13.018, 23.131). \nn
\eea
As is clearly seen from the above, the growth data contribute a $\chi^2\sim 13$ which would correspond approximately to a $\sim2.0$ or $1.8\sigma$ effect for a model with 6 or 7 parameters respectively, as is the case for the $\Lambda$CDM and $\gamma$-gravity model.

This is obviously a significant contribution to the total $\chi^2$ which clearly supports our claim that the growth data not only cannot be omitted, but they can play a crucial role in discriminating $\gamma$-gravity from $\Lambda$CDM, especially in the future with the upcoming data from the Euclid and LSST surveys.

\section{Conclusions \label{conclusions}}
In this paper we checked whether the $f(R)$ $\gamma$-gravity model is allowed by the latest observational data.
In particular, we placed constraints on the $\gamma$-gravity model by performing a joint statistical analysis using the recent cosmological data, namely SnIa (JLA), BAOs, $H(z)$, CMB shift parameters from Planck and growth rate data.

We found that $\gamma$-gravity is very efficient and in excellent agreement with observations. Furthermore, we applied the AIC criterion in order to compare the $\gamma$-gravity model with the concordance $\Lambda$CDM cosmology. We found that the value of the model pair difference is close to $\sim 2$ which suggests that the $\gamma$-gravity model is statistically equivalent with that of $\Lambda$CDM. The latter result is important because the $\gamma$-gravity model, which unlike the Hu-Sawicki \cite{Hu:2007nk} and the Starobinsky \cite{Starobinsky:2007hu} models does not include $\Lambda$CDM as a limit, can be seen as a viable alternative cosmological scenario towards explaining the accelerated expansion of the universe.

Finally, it is interesting to mention that the number count analysis based on N-body simulations~\cite{Santos:2016vdv} has shown that the predicted halo mass function is in some tension with observations. However, these theoretical predictions were assumed to be ~\cite{Santos:2016vdv} in the range $\alpha \in \[1.05,1.5\]$, while our statistical analysis provides a much higher value of $\alpha=1.892\pm 0.198$. We argue that large values of $\alpha$ could alleviate the halo-mass function problem.

\section*{Acknowledgements}
C.~\'{A}lvarez Luna acknowledges support from a JAE Intro 2017 fellowship through Grant No. JAEINT17\_EX\_0043 and the hospitality of the IFT in Madrid during the Fall semester of 2017.

S.~Basilakos acknowledges support by the Research Center for Astronomy of the Academy of
Athens in the context of the program ``Testing general relativity on cosmological scales''
(ref. number 200/872).

S.~Nesseris acknowledges support from the Research Project
FPA2015-68048-03-3P [MINECO-FEDER], the Centro de Excelencia Severo Ochoa Program
SEV-2016-0597 and from the Ram\'{o}n y Cajal program through Grant No. RYC-2014-15843.

\raggedleft
\bibliography{bibliography}
\end{document}